# Tin Titanate – the hunt for a new ferroelectric perovskite

J. Gardner,[1] Atul Thakre,[2,3] Ashok Kumar,[2,3] J. F. Scott,[1,4]

1 School of Chemistry, University of St. Andrews, St. Andrews, KY16 9ST, United Kingdom
2 CSIR-National Physical Laboratory (CSIR-NPL), Dr. K. S. Krishnan Road, Delhi 110012, India
3 Academy of Scientific and Innovative Research (AcSIR), CSIR-NPL Campus, Dr. K. S. Krishnan Road, Delhi 110012, India
4 School of Physics and Astronomy, University of St. Andrews, St. Andrews, Fife KY16 9SS, United Kingdom

Email: ashok553@nplindia.org and jfs4@st-andrews.ac.uk

Abstract

We review all the published literature and show that there is no experimental evidence for homogeneous tin titanate $SnTiO_3$ in bulk or thin-film form. Instead a combination of unrelated artefacts are easily misinterpreted. The X-ray Bragg data are contaminated by double scattering from the Si substrate, giving a strong line at the 2-theta angle exactly where perovskite $SnTiO_3$ should appear. The strong dielectric divergence near 560K is irreversible and arises from oxygen site detrapping, accompanied by Warburg/Randles interfacial anomalies. The small (4 $uC/cm^2$) apparent ferroelectric hysteresis remains in samples shown in pure $(Sn,Ti)O_2$ rutile/cassiterite, in which ferroelectricity is forbidden. Only very recent German work reveals real bulk $SnTiO_3$, but this is completely inhomogeneous, consisting of an elaborate array of stacking faults, not suitable for ferroelectric devices. Unpublished TEM data reveal an inhomogeneous SnO layered structured thin films, related to shell-core structures. The harsh conclusion is that there is a combination of unrelated artefacts masquerading as ferroelectricity in powders and ALD films; and only a trace of a second phase in Cambridge PLD data suggests any perovskite content at all. The fact that X-ray, dielectric, and hysteresis data all lead to the wrong conclusion is instructive and reminds us of earlier work on copper calcium titanate (a well-known boundary-layer capacitor).

Ferroelectrics are technologically and commercially important materials used in a wide range of applications including capacitors, memories and for applications utilising their piezoelectric and pyroelectric properties. A large proportion of commercial devices use the ubiquitous ferroelectric material lead zirconate titanate, (PZT, $PbTi_{1-x}Zr_xO_3$) due to high $T_C$, large polarisation etc.. Lead titanate and PZT with c.a. > 50 % $Ti^{4+}$ adopts a highly tetragonal perovskite, $ABO_3$ structure below $T_C$ due to the $Pb^{2+}$ lone pair. However, the deleterious effects of lead on the environment and human health have led to the search for lead-free ferroelectrics.

Tin is within group 14 of the periodic table, adjacent to lead, and like $Pb^{2+}$, $Sn^{2+}$ possesses an electron lone pair. However, while $PbTiO_3$ / PZT are well known materials, that is not the case with perovskite $SnTiO_3$. Conversely, perovskites with tin in 4+ oxidation state at the B site, eg. $BaSnO_3$, have been reported in the mid-part of the 20[th] century. [1]

Interest in perovskite $SnTiO_3$ was renewed due to recent theoretical/computational studies [2-13], with predictions this material would be a stable lead-free, room-temperature ferroelectric with high switched polarization (ca. 100 $\mu C/cm^2$), [5,8,12] capable of replacing lead zirconate-titanate (PZT) for many device applications. This has led to renewed efforts to synthesis $SnTiO_3$ in both bulk and film form [14-20].

Substitution of small amounts of tin into existing perovskite structures is possible although the site occupancy appears to differ – eg. in $(Ba,Ca)TiO_3$ $Sn^{2+}$ primarily occupies the A-site [21], yet in $SrTiO_3$ compounds occupation of both A- and B-sites is reported [22-25] . This difference may simply be related to the control of the Sn oxidation state ($Sn^{2+}$ in A-site, smaller $Sn^{4+}$ in B-site). Moreover, it is not impossible to form other homogeneous stable $Sn^{2+}$ oxides including tungstates and titanates, as shown by the work on $Sn_2TiO_4$ [26] and also $SnNb_2O_6$ [27], $SnTa_2O_6$ [28] and $SnWO_4$. [29]

Due to the predicated attributes perovskite structured $SnTiO_3$ numerous methodologies have been employed over a number of years due to its potential as a lead free device material, with limited success. Traditional solid state methodologies employing high temperature reactions due to the issues caused by the disproportionation of SnO and oxidation of the $Sn^{2+}$. [30] Reports of such solid state reactions are largely absent from the published literature due to this lack of success. A Bachelor's thesis by Gubb has shown a number of possible synthetic routes, including solid state and sol-gel reactions performed under various atmospheres and at a wide range of temperatures.[18] No perovskite structured material was formed, although a rutile structured SnTiO$_4$ [(Sn,Ti)O$_2$] was formed using aqueous methods and low temperature reactions.

Most recent work has focused primarily on using soft chemistry routes, such as sol-gel or co-precipitation, and thin-film techniques. In this review we examine the current literature reporting experimental studies of SnTiO$_3$.

**Co-precipitation from chlorides (Kerala)**

Conventional solid state methods employing high sintering temperatures typically lead to oxidation/disproportionation of Sn$^{4+}$ precursor reagents. Attempts to prevent the oxidation of the Sn$^{2+}$ have included "soft" chemistry preparation routes which typically permit lower sintering temperatures and superior admixing compared with traditional solid state methods. Two papers report attempts to produce samples of SnTiO$_3$ via a co-precipitation method using SnCl$_2$ and TiCl$_4$ as starting materials [14, 15].

The first, [14] refers to "perovskite type tin titanate (SnTiO$_3$)" in the introduction while discussing its proposed properties; however, while the author discusses "tin titanate" and refers to "SnTiO$_3$ powders", the structural data reported are sparse. The paper states that "the formation of tin titanate is evidenced by the characteristic peak of tin titanate in the XRD spectrum" and lists the positions of three peaks from a powder X-ray diffraction (PXRD) pattern measured from powders calcined at 850 °C. However, no PXRD reference pattern or crystallographic model is referenced, and all other structural/crystallographic information is absent – symmetry (space group and point group), lattice parameters, atom positions, etc. are not reported.

The positions of the three peaks (31.8°, 33.5° and 39.8° 2θ) are similar to those of a paper with a simulated perovskite pattern published by Zhao *et al*. [19], based on calculations from previous theoretical studies [5, 12]. This would suggest a large tetragonal strain (cell of $a \approx b \approx 3.80$ Å and $c \approx 4.09$ Å from [19]. However, other tin-titanium containing oxides possess PXRD peaks close to these 2θ positions listed (e.g., simulated PXRD patterns from crystallographic models from the Inorganic Crystal Structure Database (ICSD) - collection codes 161282, 410690, 30664 [31-33]).

As bulk perovskite structured SnTiO$_3$ would be a novel and commercially important material, it is difficult to conclude that ref [14] successfully synthesised the material, due to the lack of structural data presented. Additionally, no dielectric or ferroelectric properties were studied.

A subsequent paper by a different group used a similar co-precipitation method [15] and unambiguously describe the resulting compound as perovskite. Greater details of structural studies are provided; however, the paper is somewhat contradictory: They describe the formation of a new phase and list PXRD peak positions (21.62, 29.01, 33.4, 34.8, 37.8, 49.09, 50.96 and 59.8° = 2θ) cited from an abstract of a master's thesis from which the peak positions were obtained. [34] The thesis abstract, (which appears to be available only via a dissertation publishing site) states that the PXRD peaks of their material, obtained from a different synthetic route, was difficult to match to a Powder Diffraction File card (a reference pattern from database). The authors of Ref [15] then cite Ref [14] when describing their PXRD data as having "characteristic peroveskite (*sic*) structural values", however, other than the peak at 33.4° the other two values from Ref. [14] do not match those listed above. Furthermore, there are several issues with the accompanying PRRD pattern: there is a greater number of peaks than those listed in the text; there appear to be two different sets of peaks with different peak broadness overlying each other (it is not stated if one set is a reference pattern, and information on the symbols over certain peaks is likewise omitted), and the figure image is blurred to an extent that the axis scales are illegible).

Therefore, based upon the early work in India by itself, one cannot definitively conclude that perovskite SnTiO$_3$ was synthesised.

**PLD films (Cambridge)**

The deposition of materials in the form of thin films has various technological advantages, however, has also enabled materials to be formed which are unstable in the bulk form. As such, much recent work on tin titanate has concentrated on the fabrication of thin films using a number of deposition techniques.

Fix *et al*. produced films deposited using pulsed laser deposition (PLD) on a number of different perovskite structured substrates, the unit cell of which possessed lattice parameters close to desired phase [16]. One of the disadvantages of film deposition is the typically limited information using standard laboratory diffraction (i.e. x-ray diffraction, XRD) techniques as films are typically thin and often orientated such that only a small number of crystallographic planes satisfy diffraction conditions. This results in only a small number of often weak peaks being observed. As such, only one line is observed in their diffraction data, Fig. 1, which is somewhat inconclusive. However, electron diffraction data, Figure 2, shows a triclinic cell consistent with an ilmenite $SnTiO_3$ structure with $\alpha = 90°$, $\beta = 90°$, $\gamma = 120°$, and lattice dimensions of $a = 5.04$ Å, $b = 5.19$ Å, $c = 14.56$ Å. In discussing why the ilmenite structure is formed Fix et al. note that, while a number of theoretical studies predict a perovskite $SnTiO3$ structure, Ref [35] calculates that the ilmenite structure is more stable in a reducing atmosphere.

The PLD films made by Fix et al do not exhibit any indication of ferroelectricity under measured conditions; however, additional weak lines were detected in reciprocal XRD space maps, which do not originate from the substrate or ilmenite phase. The authors propose that this minor may originate from very small quantities of perovskite, with inferred unit cell of $a = b = 3.76$ Å and $c = 4.18$ Å, similar to those previously predicted theoretically for perovskite $SnTiO_3$. [12, 16] We note that Pitike et al.[9] conclude that SnTiO3 is stable in P4mm structure only for <a,b> < 3.87 Å and distorts to monoclinic Cm (C1m1) for larger lattice constants.

Using a tetragonal perovskite structural model, we have simulated an XRD pattern, Figure 3, of this phase to determine the 2θ positions of the Bragg peaks as they would appear with Cu K$\alpha_1$ radiation. The perovskite structural model used was modified from a $PbTiO_3$ structure from the Inorganic Crystal Structure Database (collection code 61168, [36]) with *P4mm* symmetry as this symmetry has previously been used in computational/theoretical studies. [2, 37] $Pb^{2+}$ was replaced by $Sn^{2+}$ at the A-site and the lattice parameters were altered to those reported by Fix *et al*. – no other changes were made to e.g. atom positions. The three peak positions reported by Beenakumari in the powder chloride synthesis work discussed above [14], are close to those in the simulated pattern. Oxides of tin and titanium and metallic tin also have peaks close to those three 2θ values and the powder in [14] had been heated to 850 °C at which $Sn^{2+}$ compounds usually decompose/disproportionate [30] Therefore, while this is not enough to conclude that perovskite $SnTiO_3$ was actually made in Ref [14], it does demonstrate that the values are consistent with the minor perovskite phase present in the PLD films.

- PLD films form phase consistent with ilmenite structure.
- Minor phase which may be perovskite $SnTiO_3$.
- No switchable polarisation from P-E measurements

**YMnO$_3$/SnTiO$_{3+x}$ heterostructure PLD films**

Zhao and co-workers attempted to stabilise tin titanate by growing multilayer PLD films with alternative layers of the hexagonal YMnO$_3$ to produce YMnO$_3$/SnTiO$_{3+x}$ with varying numbers of layers [19]. These films were deposited on (100), (111) and (110) Nb:SrTiO$_3$ (Nb:STO) substrates with 2, 4 or 8 layers. Zhao et al assigned XRD peaks from these films, Figure 4, as follows: ilmenite [two layer film deposited on (100) Nb:STO], a mix of perovskite and ilmenite [4 layer films on (100) and (110)], perovskite and an unidentified phase [4 layers on (111)] and perovskite [8 layers on (100)]. As seen in Figure 4, with the exception of a peak at approximately 33° two theta, the presence and appearance the other peaks assigned to a perovskite structure changes with layer and substrate orientation and while this may be due to the orientation of the film, the authors do not discuss this or any possible origin (e.g. epitaxy), They do however note the need for further structural characterisation using TEM and state their intension to carry this out, however, no further papers appear to have been published to date.

P-E data, Fig. 2 in Ref [19], on the 2 layer film (ilmenite XRD) shows a very small measured polarisation of < 0.05 μC/cm$^2$ which would be consistent with the non-polar FeTiO$_3$ ilmenite structure they assign and with the linear P-E response (*i.e.* not ferroelectric) Fix *et al.* obtained from their principally ilmenite PLD film. [16] For films with four or more layers they measure a larger polarisation but somewhat lossy P-E loops and state that "contributions to the hysteresis loop from movable charges was significant" in certain films. Charge injection is one of a number of artefacts found in bulk and thin films which may obscure the true properties of a material. [38-41] PFM data also indicates possible switching, however, as seen later in recent PFM work on ALD films possible charging effects may be misleading. Without PUND measurements and additional PFM experiments, some ambiguity remains. Additionally, the contribution from the intrinsic magnetic and ferroelectric properties of the YMnO$_3$ layers or from interfacial effects it is difficult to establish the properties of the SnTiO$_{3+x}$ components of the film structure.

- XRD pattern consistent with perovskite and ilmenite structures
- No microscopic study of film or detailed structural investigation using electron diffraction
- Possible ferroelectric switching, however, further experiments would be required to exclude artefacts and influence of YMnO$_3$ layers.

**ALD film**

Chang and Selvaraj [42] reported the atomic layer deposition (ALD) of tin and titanium precursors (TDEAT and Sn(acac)$_2$ yielding films of tin titanate (SnTiO$_x$) on Si(100) substrates with the aim of producing perovskite structured SnTiO$_3$. This synthesis method typically uses lower reaction temperatures and control of oxidation states and composition is possible by variation of the annealing conditions of the deposited film [43-46]. Deposited films were annealed under different atmospheres (O$_2$, N$_2$ and H$_2$) at 350 and 650 °C for 5 mins. They found anatase appeared to be the major phase present, on both as-deposited and annealed films, using grazing incident angle X-ray diffraction (GIXRD). This was based on assigning two broad peaks at c.a. 55 degrees 2θ to the 105 and 211 reflections from the anatase structure. However, piezoelectric switching hysteresis loops were demonstrated on films annealed at 350 °C under O$_2$,[42,43] which led to further studies on the film [A2], including polarisation-electric field (P-E) measurements, dielectric and further PXRD data.

Polarisation-electric field (P-E) measurements yielded hysteresis loops with 3μC/cm$^2$ remnant polarization at room temperature and further PUND data was used to determine the existence of switched polarisation in the film [17]. The latter measurement uses a different pulse sequence from typical dynamic P-E measurements, consisting of two successive electrical pulses of the same polarity. Assuming the ferroelectric material is switched during the initial pulse, the second pulse should have no contribution originating from the switching of ferroelectric domains and therefore reveal contributions from extrinsic sources such as leakage current. [47] This is, however, not always the case due to e.g. relaxation.

A frequency dependent peak was observed in the temperature dependent relative permittivity with T$_m$ ≈ 450 – 600 K.  Additional peaks were observed in PXRD, although the data are somewhat complicated by the presence of a nominally forbidden Si(200) peak overlying a peak at 33° 2θ attributed to the film. This substrate peak is basis-forbidden but may be present in diffraction patterns due to multiple (double) diffraction (Umweganregung) [48, 49]. Bragg peaks in PXRD patterns originate from the diffraction of incident X-rays from lattice planes of a crystalline material. In the case where nominally forbidden Bragg peaks are observed due to multiple diffraction, the incident X-rays have previously diffracted from a lattice plane rather than originating directly from the X-ray source.  The forbidden peak may have variable intensity and exhibit shoulders and sub-peaks at about 33 ±2.5° = 2Θ, as shown in Figure 1 of Ref [48]. PXRD patterns of the substrate lack any such shoulder, Figure 5.

Hwang has shown that for PXRD diffractometers operating in Bragg-Brentano geometry, the intensity of this forbidden peak varies with Φ, the angle normal to the plane surface of the sample [49]. On such diffractometers it is common to spin the sample within this axis to improve the quality of the data collected, which leads to an averaging of the diffraction from all Φ angles. New data, Figure 5, on one of the ALD films were collected during which the sample was not spun, and as can be seen both the sharp forbidden Si peak and that assigned to the perovskite film, are absent. A peak at high angle (c.a. 117 ° = 2θ) also coincides with a possible Si peak, leaving only the peak around 55° = 2θ which consistently occurs in conventional XRD and GIXRD; therefore the structure of the film as determined by X-ray diffraction is somewhat ambiguous. Hence, despite significant improvements over the earlier Kerala data, the XRD results are by themselves inconclusive.

AFM/PFM

AFM/PFM has thus far failed to reveal any ferroelectric domains in these samples or in the ALD films studied previously. There are several possible reasons: (a) Any domains are expected to be very small (<30 nm), since the grains are very small, where previously measured; (b) the samples almost certainly are heterogeneous, and the outermost surface layer is ca. 15 nm thick, according to TEM; (c) the surfaces do not appear to be extremely clean, and attempts to clean them may destroy the films. Strong surface charging of half-surfaces occurs, Figure 6; however, as no domains are visible, this may just be surface charging of the whole sample.

In attempt to determine if charge writing is being observed Electrical force microscopy was performed. Two surface areas of the ALD films (indicated by boxes in Figure 6) with opposite polarities were written and then Electrical force microscopy preformed to probe surface charging. EFM signal is usually indicative of charge being injected into surfaces for myriad of reasons. As seen in Fig. 6, the signature of charge writing in the written boxes is evident. This signature stays for a few scans meaning it could explain the signal decay seen in piezoresponse images, Fig. 7. The result still does not rule out ferroelectricity completely but does explain the lack of retention of the written states.

**Sol-gel**

In an attempt to reconcile some of the results discussed above, tin titanate (SnTiO$_3$) nanoparticles were most recently synthesized in Delhi synthesized using a sol-gel method. Tin acetate and titanium (IV) isopropoxide were employed as precursors. Methanol, acetic acid (glacial) and 2-methoxy ethanol were used as the solvents. Initially in the nanoparticle preparation process, a calculated amount of tin acetate was dissolved in methanol (say sol 'A'). Secondly, a solution of the acetic acid (glacial) and 2-methoxy ethanol in a 1:1 ratio was prepared; followed by dissolving the calculated amount of titanium isopropoxide into it (say sol 'B'). The sol 'A' and sol 'B' were then instantly mixed and stirred for half an hour. The prepared solution was then heated at 250 ºC on hot-plate for 24 hrs. The obtained nanoparticles were calcined at 550 ºC for 4 hrs. The circular pellets having diameter and thickness 1 cm and 650 µm respectively, were pressed. The prepared pellets were sintered at 550 ºC for 4 hrs. In order to carry out the electrical measurements, the electrodes of In-Ga alloy were attached on top and bottom of the pellet.

PXRD data of a new tin titanate powder synthesised by sol-gel are shown in Figure 8. Bragg peaks exhibit extensive broadening, likely size-broadening due to the small size of the crystallites (determined to be c.a. 8-10 nm from TEM). Peak positions are consistent with a rutile structure and Rietveld refinement using a rutile structural model (International crystallographic database collection code – 161282) accounts for all peaks and gives a reasonably good fit (goodness of fit parameters $\chi^2$ = 1.4, Rwp = 0.129,  R(F2) = 0.12). The crystallographic model from the database originated from ref [50] for a material with composition $Sn_{0.39}Ti_{0.61}O_2$, using tetragonal space group $P4_2/m\,n\,m$). The refined unit cell has lattice parameters of $a = b = 4.67$   $c = 3.08$ Å, which is somewhat intermediate of rutile structured TiO$_2$ and SnO$_2$ (cassiterite). [51] This is unsurprising due to the Sn$^{4+}$ cation being larger than Ti$^{4+}$ [52]. It should be noted that PXRD is a bulk technique which obtains the spatially averaged structure and that Ti-Sn segregation can occur [53].

SEM and TEM data, Figures 9 and 10 show nano-crystallites with homogenous morphology. SEM/EDAX gives a 1.1:1 Sn:Ti ratio. This ratio is consistent with both rutile and perovskite structures. HRTEM data also gives 110 planes with d-spacing of 0.33 nm

Raman

The odd-parity zone-center vibrations in the cubic phase have been calculated as [ ]: TO(1), soft (imaginary frequency); LO(1), 80 cm-1; TO(2), 126 cm-1; LO(2), 375 cm-1; TO(3), 505, cm-1; and LO(3), 689 cm-1. A few experimental values have been reported [7] for wave numbers above 400 cm-1, but they are not in close agreement with predictions, with phonons at 455 cm-1, 619 cm-1, and slightly less than 800 cm-1. As expected, these are close to those in rutile $TiO_2$ or rutile-structure cassiterite $SnO_2$ (the latter at 474, 632, and 774 cm-1). [ , ]

Raman data were obtained for the ferroelectric phase in the present work (Table 1 and Fig. 11) but symmetries are uncertain. Even if there were comparable amounts of rutile-structure and perovskite, the Raman spectra of the latter would be much weaker because in the cubic perovskite phase first-order Raman scattering is forbidden (all ions at inversion centres); so the spectra might be dominated by cassiterite/rutile.

Table 1. Phonon frequencies in tin titanate

| Calculated (present work) | Symmetry phase | Experiment | LO Cubic $SnO_2$ | Rutile $TiO_2$ | Cassiterite cm$^{-1}$ |
|---|---|---|---|---|---|
| **90** | E(TO) | 100 | | | |
| **134** | E(LO) | 140 | 80 | 143 $B_{1g}$ | 123 $B_{1g}$ |
| **218** | E(TO) | --- | | | |
| **230** | E(LO) | --- | | | |
| **232** | A1(TO) | --- | | | |
| **243** | A1(LO) | 248 | | | |
| **319** | E(TO) | --- | | | |
| **399** | A1(TO) | --- | | | |
| **448** | E(LO) | 442 | 375 | 443 $E_g$ | 474 $E_g$ |
| **462** | A1(LO) | --- | | | |
| **530** | E(TO) | --- | | | |
| **633** | E(LO) | 628 | | 610 $A_{1g}$ | 632 $A_{1g}$ |
| **704** | A1(TO) | --- | | | |
| **820** | A1(LO) | 798 | 689 | 844 $B_{2g}$ | 774 $B_{2g}$ |

Electrical data

Dielectric measurements as a function of temperature are presented in Fig. 12. Frequency-dependent features in both the relative permittivity and normalised loss, tan δ, occur at similar temperatures to that of the dielectric peak observed in the ALD film [17]. However, these features are only observed on heating – they are largely absent on cooling, Fig. 12(c) and (d) and during further heating and cooling cycles. The samples were sintered at 550 °C for 4 hours and so the observed behaviour is unlikely to originate from a transformation to a new phase.

Unfortunately, it is well established that non-ferroelectric materials deposited onto semiconducting or metal blocking electrodes can sometimes produce large dielectric peaks in the 200-500 C temperature regime which typically exhibit a frequency dependence [54]. Replacement of the Schottky electrodes Ag or Au with more ohmic contacts such as In-Ga also provides discriminating between intrinsic dielectric peaks and electrode contributions. However, transition metal doped $SnO_2$ has been reported to shows non-ohmic grain boundaries [55]

Impedance data is also similarly complex - the impedance complex plane (Cole-Cole) plot measured at 475 K shown in Figure 13 gives what appears to be a semi-circle and a low frequency tail, indicative of a Warburg element [55, 56]. This resembles the response obtained from a Randles circuit, an equivalent circuit which is often used to model processes at interfaces [55, 56]. Not only is it extremely linear over a long range, but it makes the required angle of 45 degrees (gradient of slope is 1.00, Figure 15). Equivalent circuit models should not be derived based on single data sets, however, as often other electroactive regions within the sample being tested simply do not contribute to the data at that specific temperature (i.e. their time constant is outside the spectroscopic measuring window). With increasing temperature the low frequency tail becomes non-linear before forming another semi-circle. This higher frequency section of this new feature is still somewhat linear initially (Figure 15) – it is therefore difficult to determine if the 'tail' represents true diffusion, as Warbug elements are normally interpreted, or merely the high frequency response of the other feature moving into the measured frequencies. If it is the former, such linearity has been reported, attributed to a change from semi-infinite to finite length diffusion [56] with the linear region being described as the "transition region" [57].

There is also a large and as yet unexplained, change in sample resistance of x100 We attribute all these phenomena in dielectric response near T=500K, including the "Warburg" element as due to abrupt emptying of oxygen traps. Artefacts from e.g. the electrode-interface effects and oxygen vacancy de-trapping etc. are common [58], and the observed features in dielectric and impedance data are likely to be extrinsic in origin. The inference of oxygen site depinning is not just an unsupported hypothesis. The reversible surface charging (Fig.6 and 7) reported by Amit Kumar at Belfast in his AFM study (bright metallic with +10V applied; dark with -10V applied) is compatible with that, as is his lack of domain walls in the AFM images. And equally important, Beenakumari reported an ambient activation energy of 1.08 eV for her initial samples; this energy of 0.9-1.1 eV is almost always attributed to oxygen vacancies in pseudo-perovskite oxides. [59]

(Sn,Ti)$O_2$ compounds have been studied for use as a potential device material for gas sensors and varistors [60] various challenges. These issues include immiscibility of tin/titanium, [60], reported different conduction processes - band conduction in $SnO_2$ and hopping in $TiO_2$ [61] and differences in conduction with thermal cycling [62]

P-E measurements show what appears to be a leakage contribution at lower frequencies; however, at 100 Hz appear to be those of a linear (albeit somewhat slightly leaky) dielectric, Figure 15 (a). These hysteresis loops are reproducible but cannot be intrinsic, since they do not evidence a Curie temperature above which they disappear. Even more important, they are present for specimens that are shown to be pure rutile/cassiterite structure. They may arise from charge trapped at the electrode-dielectric interface. As discussed elsewhere in the dielectric section, we suspect they are oxygen traps.

PUND data (measurement described above [47]) of these samples indicate little or no actual switching of polarisation, Fig 15(b). The value of polarisation for successive pairs of pulses (P and U; N and D) are very similar which suggests that the primary switching pulses (P and N) do not contain a contribution from the switching of domains – i.e. that the response is purely extrinsic.

**New Ilmenite Paper**

Most recently a paper by Diehl et al. successfully used soft chemistry to produce a bulk form of SnTiO$_3$. [20] To avoid oxidation/disproportionation they used a multi-step metathesis synthesis: first forming a layered K$_2$Ti$_2$O$_5$ precursor which was heated with SnCl$_2$.2H$_2$O at 300 °C under vacuum for 24 hours, after which the KCl salt by-product was removed with washing, leaving SnTiO$_3$.

Diehl *et al*. detected no perovskite, finding instead that their bulk SnTiO$_3$ powder formed an ilmenite-like structure, however, they were not able to match the indexing directly with the previously reported ilmenite structured PLD film.[16] They determine that Ilmenite-like layers were present, separated by van der Waals gaps from the lone pairs of Sn$^{2+}$, which form multiple stacking orders and twin domains [20]. A combination of a number of polytypes, each modelling different stacking sequences, was required to account for the observed structural data with Electron energy loss spectroscopy and nuclear magnetic resonance data supporting the resulting local structure/symmetry.

Despite the structural complexity introduced by the different stacking sequences, the bulk SnTiO3 obtained by this group shares similarities most with the PLD films made by Fix et al. [16] – i.e. major product is ilmenite or ilmenite-like structure rather than perovskite. The non-layered nature observed in the PLD films may simply be a be a consequence of the synthesis method. Diehl et al. calculated that the different stacking sequences were comparable in energy in the bulk form and that, using electron localisation function and density functional theory, the lone pairs of the Sn$^{2+}$ directed the structure. [20] In the PLD films strain, low dimensionality, synthesis temperatures etc. seems to favour the simpler illmaite structure. Another possible explanation is the role of the layered K$_2$Ti$_2$O$_5$ starting material.

**Conclusions**

The synthesis of bulk perovskite SnTiO$_3$ presents many problems – the thermal stability of the SnO and propensity of Sn$^{2+}$ to oxidation. Additionally, Sn$^{4+}$ in B-site of perovskites and related structures inhibit ferroelectricity [63] Preparation routes which preserve (or generate) Sn$^{2+}$ are needed to both make the perovskite structure and for optimal ferroelectric. Delicate balance between synthesis temperatures, strain, control of atmosphere. This has most successful been achieved via the creation of PLD films, which are more difficult to characterise than bulk material, and more recently by bulk reaction under vacuum. We suggest future studies into the synthesis method of the Diehl et al. including possibilities of changing reaction conditions or potassium intermediate to influence the stacking arrangements of ilmenite and possibility of forming perovskite $SnTiO_3$. Other synthesis methods may also be used, e.g. high pressure methods. Randall et al. have recently reported the densification of SnO using cold sintering, [64] having previously used the synthesis method to produce PZT and $LiFePO_4$. [65] The low temperatures used may allow the preservation of the $Sn^{2+}$ state but favour the perovskite form.

**Overall conclusions of this review:**

Perovskite ferroelectric tin titanate has not yet been made in thin films or bulk as single-phase material. It clearly does exist, for example in a minor phase prepared via PLD, and possibly as a single-phase powder via titanium chloride processing. Despite its toxicity, this process should be examined further in laboratories with proper safety facilities.

The electrical data are largely dominated near T=500K by abrupt emptying of oxygen traps, not a ferroelectric phase transition. AFM shows that these surfaces can be emptied or filled via +10V and −10V but that domains do not form.

The PUND switching data are in themselves not unambiguous evidence of ferroelectricity.

Several studies suggest highly inhomogeneous products, with nano-phase separation. This may account for differences in published XRD data.

As a Key Issues Review, this report is not intended to be the definitive story, but merely a guide to help new researchers through the literature, which is not all self-consistent.

Acknowledgements:

The synthesized materials at CSIR-NPL and related data may be requested to email: ashok553@nplindia.org. Amit Kumar for new PFM work on ALD films. Finlay Morrison. This work was supported in St Andrews by the Engineering and Physical Sciences Research Council (EPSRC) Grant No. EP/P024637/1

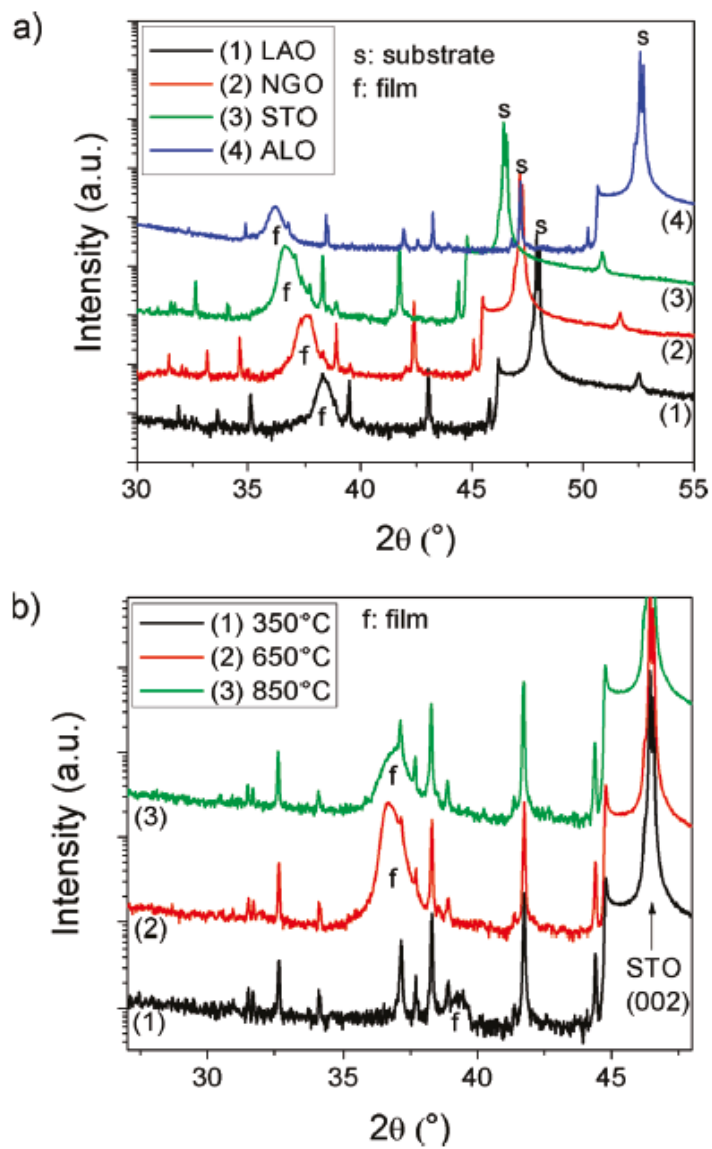

Figure 1: Powder X-ray diffraction patterns of SnTiO$_3$ films grown by Driscoll et al. showing (a) films deposited on a range of substrates and (b) films deposited on STO (001) at various temperatures. Reprinted with permission from Fix et al. Cryst. Growth Des. 2011, 11, 1422. [16] Copyright 2011 American Chemical Society.

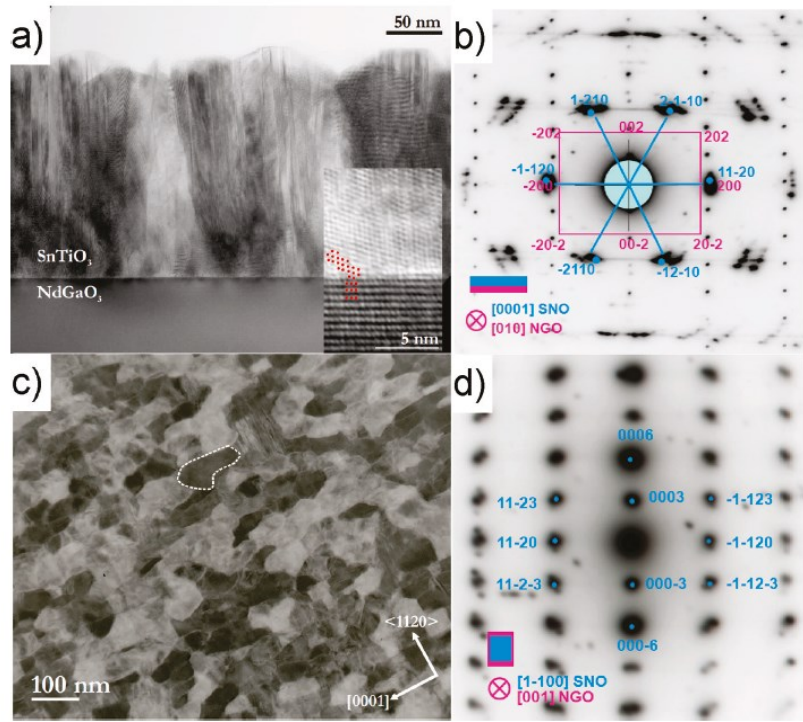

Figure 2: SnTiO₃ films grown by Driscoll et al. showing: (a) HRTEM image of film cross section, (b) diffraction pattern of with NGO [010] zone axis (hexagonal indexing), (c) TEM image (plan-view) of film and (d) diffraction pattern of with NGO [010] zone axis with hexagonal indexing (plan-view). Reprinted with permission from Fix et al. Cryst. Growth Des. 2011, 11, 1422. [16] Copyright 2011 American Chemical Society.

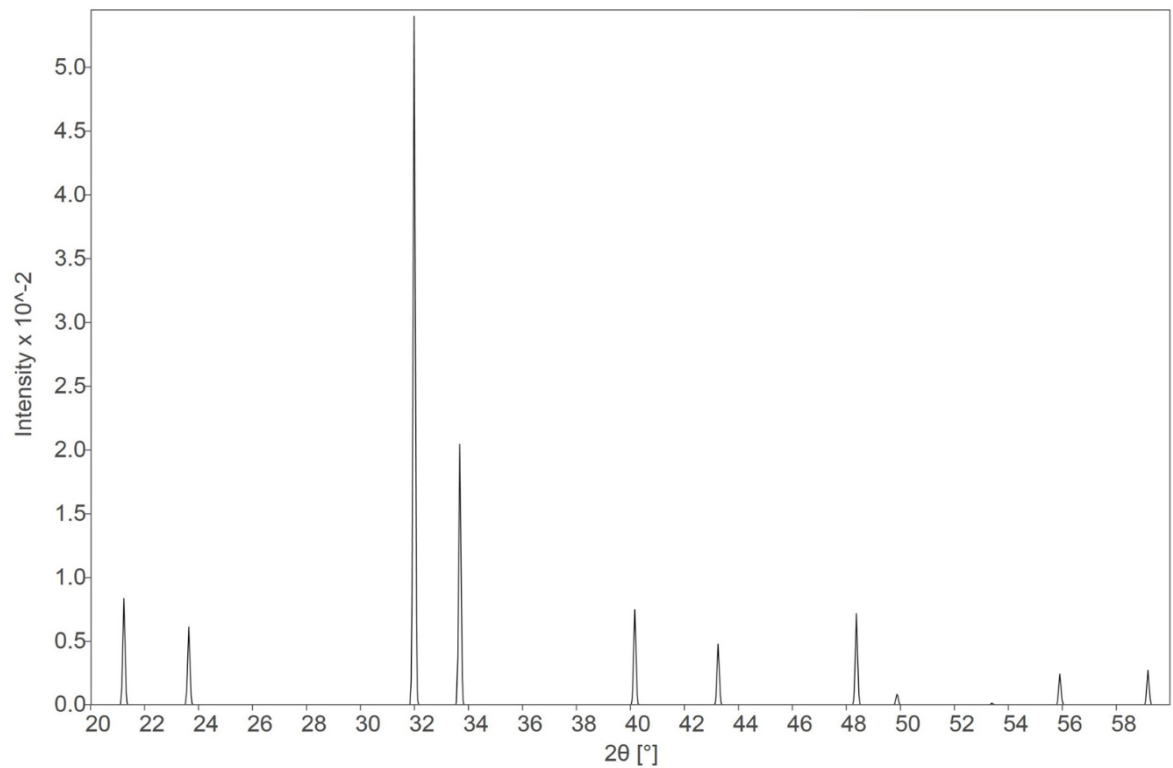

Figure 3: simulated PXRD pattern of perovskite structure (Cu Kα$_1$ radiation) using perovskite structural model (space group *P4mm*) and lattice parameters of possible minor perovskite phase from Fix *et al*. [16].

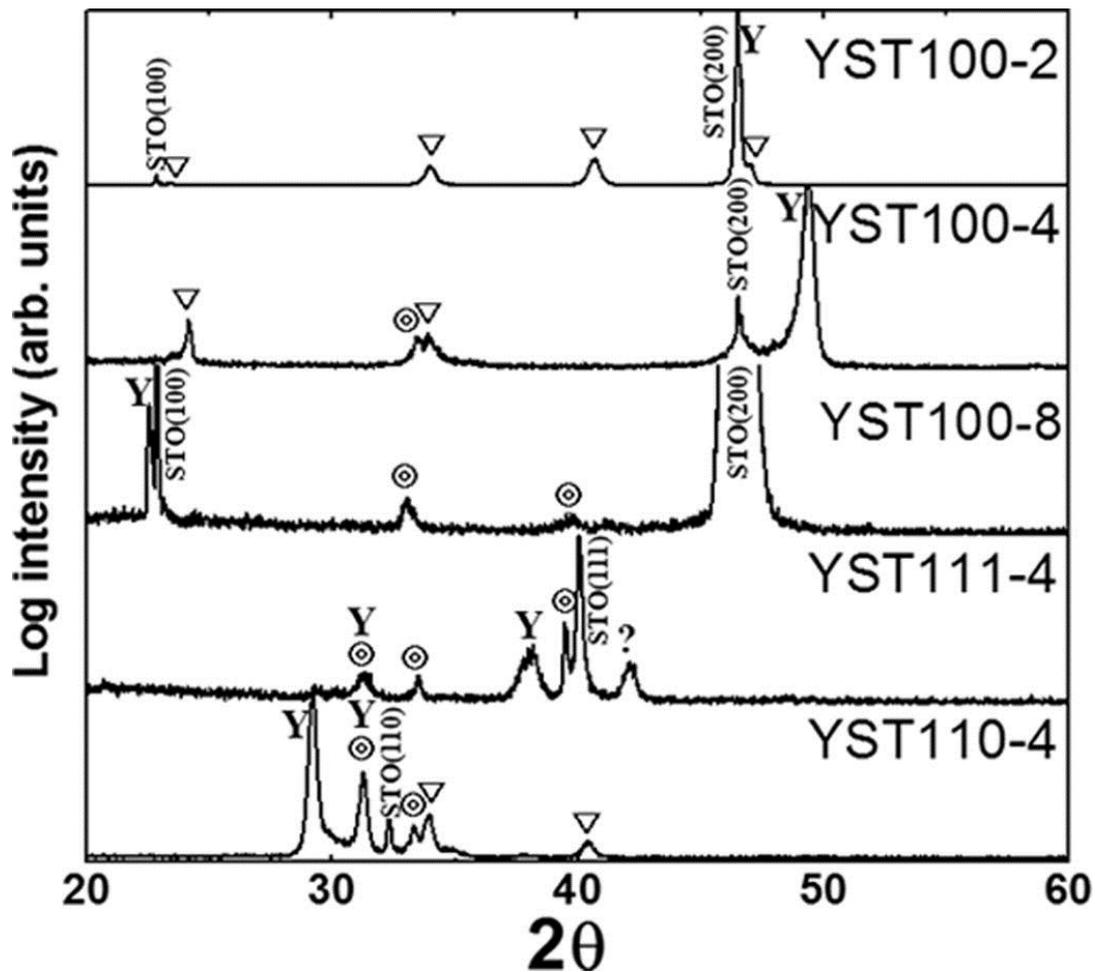

Figure 4: XRD of YMnO3/SnTiO$_{3+x}$ PLD films produced by Zhao et al. Films were deposited on 100, 111 or 110 orientated substrates with 4 or 8 layers - indicated in text on right of figure. Y = YMnO$_3$ phase, circle = perovskite, and triangle = ilmenite. Reprinted from "A new multiferroic heterostructure of YMnO$_3$/SnTiO$_{3+x}$", H Zhao *et al.*, Scripta Materialia 65 (2011) 618, Copyright (2011), with permission from Elsevier and Acta Materialia Inc.

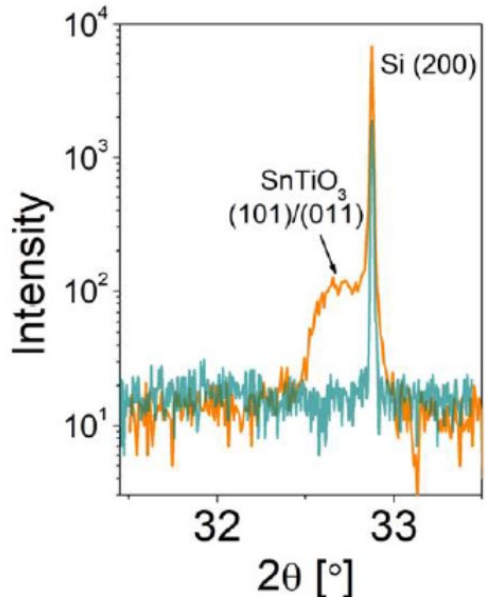

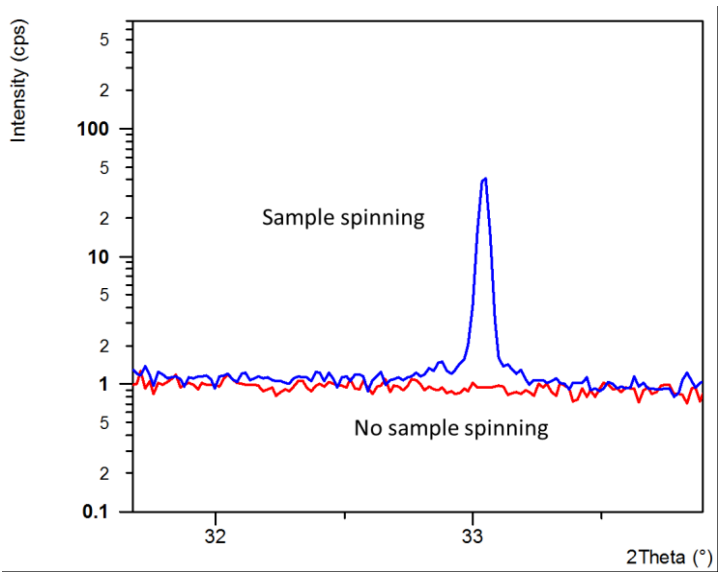

Figure 5: Powder X-ray diffraction patterns of: (a) Si(100) substrate (grey?green?) and deposited ALD film on substrate (orange), with the forbidden Si(200) and overlying peak, proposed to originate from perovskite SnTiO3, indicated (Reproduced from Ref [17], R. Agarwal, et al., Phys. Rev. B 2018, 97, 054109; (b) loss of forbidden Si(200) peak when sample sample is not spun.

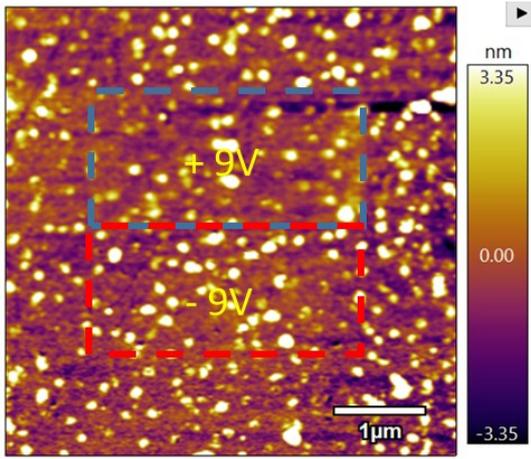
Topography

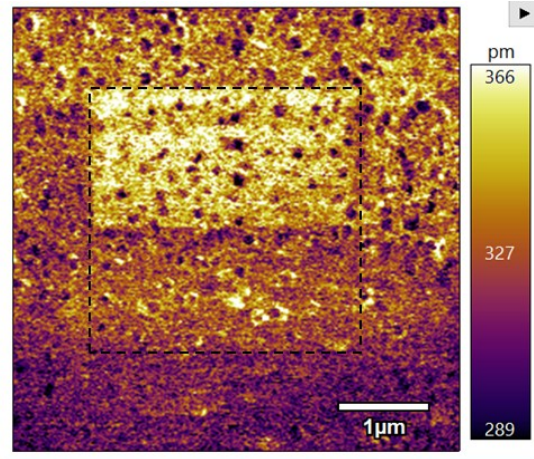
PFM amplitude

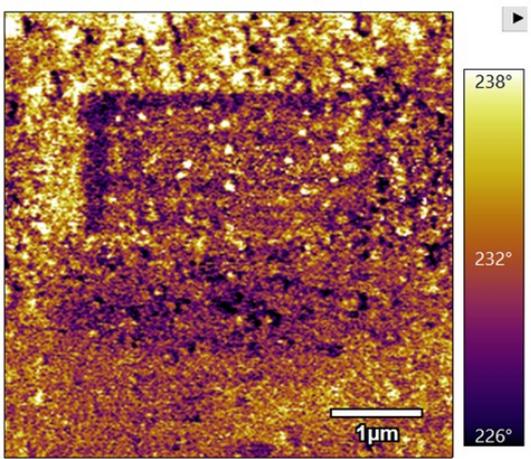
PFM Phase

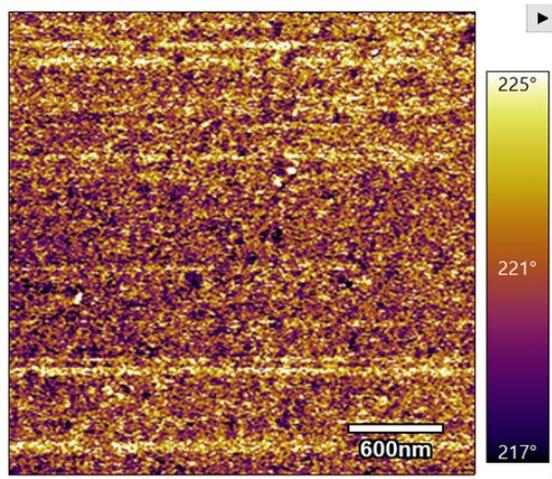
PFM Phase of inside box before biasing

Figure 6: Piezoresponse force microscopy of ALD film. Contrast in phase and amplitude show contrast suggestive of bias writing.

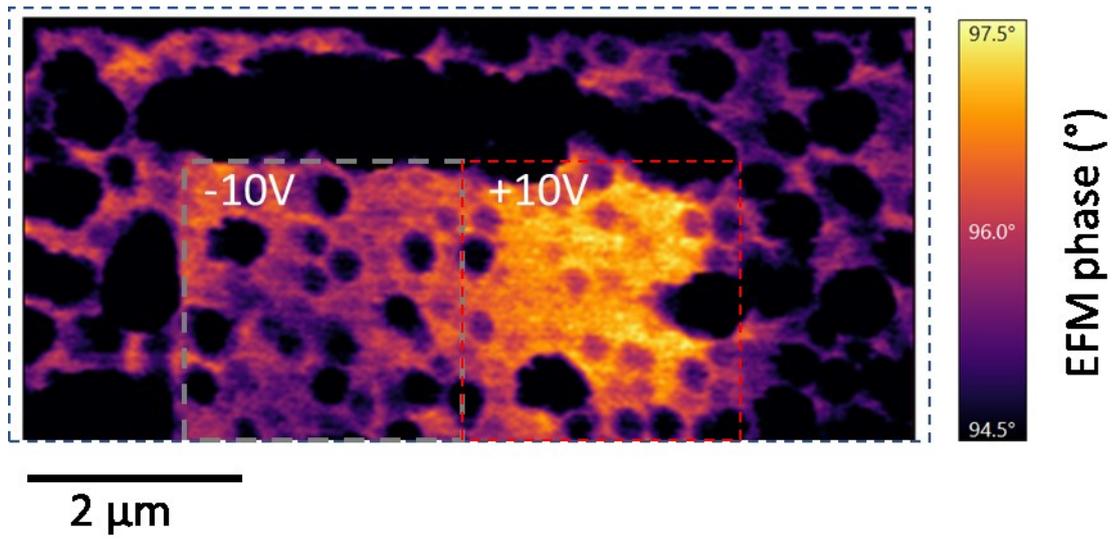

Figure 7: Electrical force microscopy of ALD film.

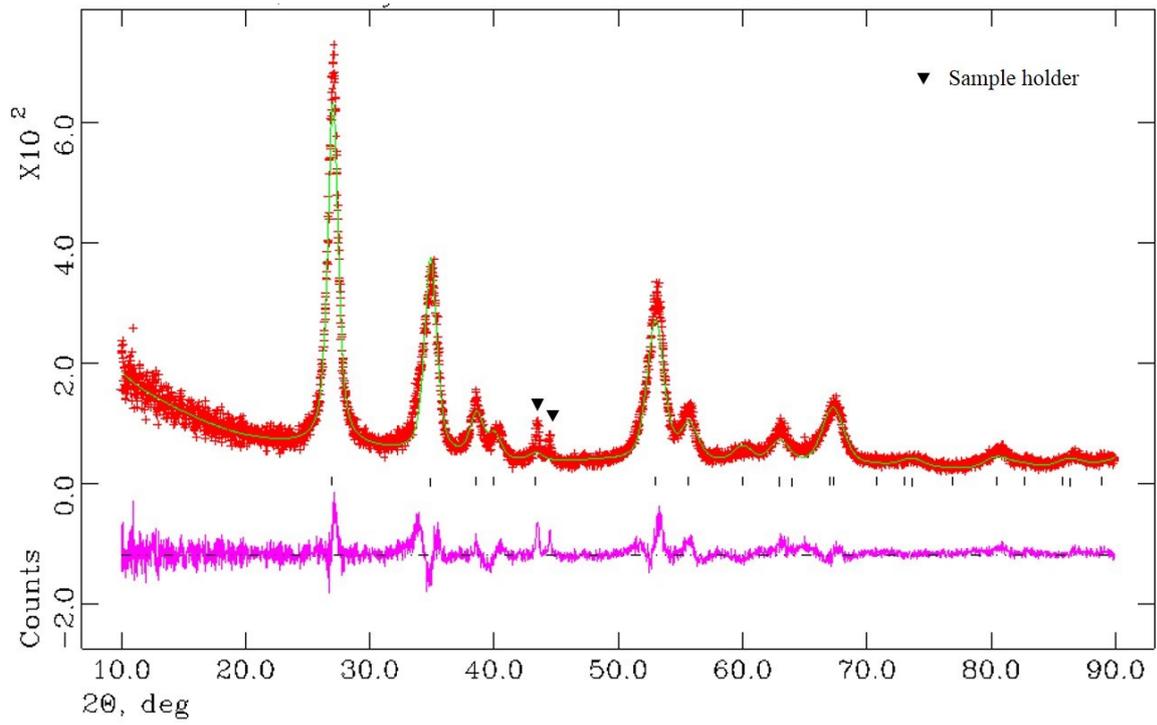

Figure 8: powder x-ray diffraction pattern of new sol-gel tin titanate. Bragg peaks consistent with rutile structure.

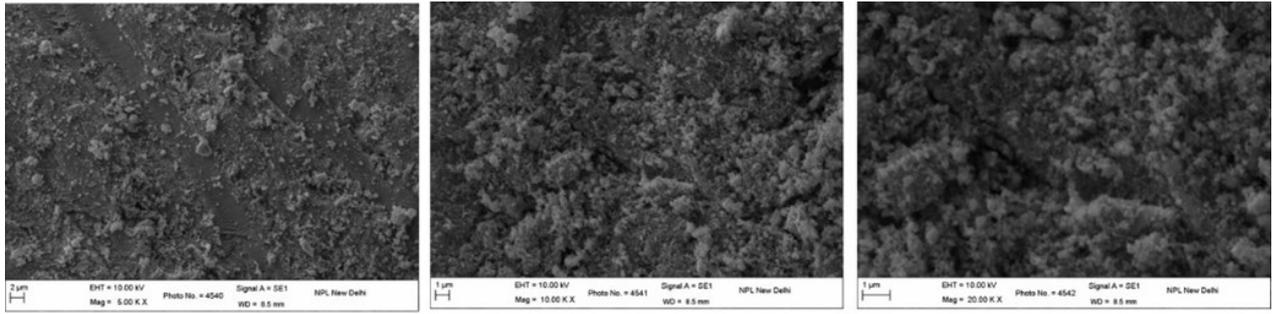
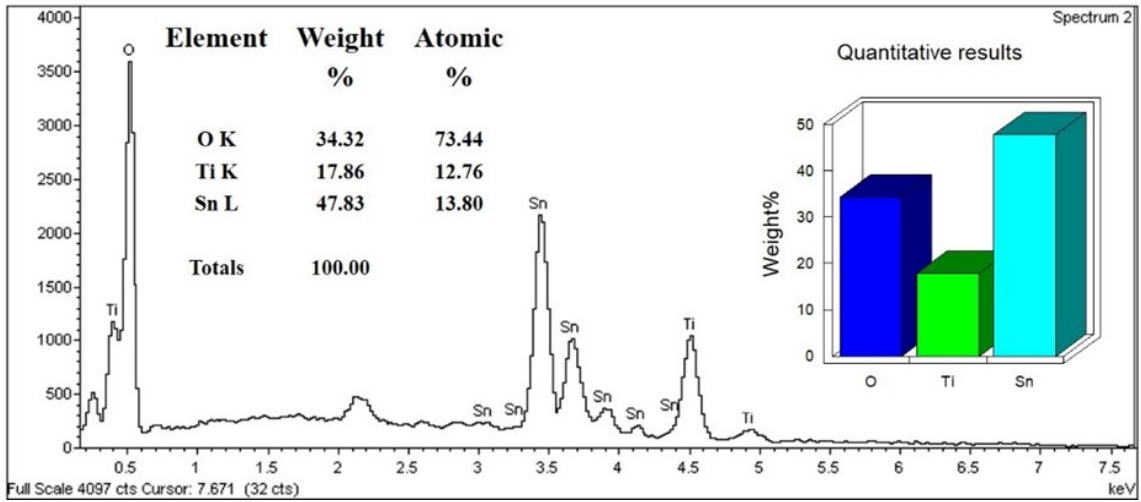

Fig. 9: illustrates the SEM/EDAX data, which demonstrate a 1.1:1 ratio of Sn/Ti in the new sol-gel samples, with 12.8% (atomic) Sn and 13.8% Ti and 73% oxygen.

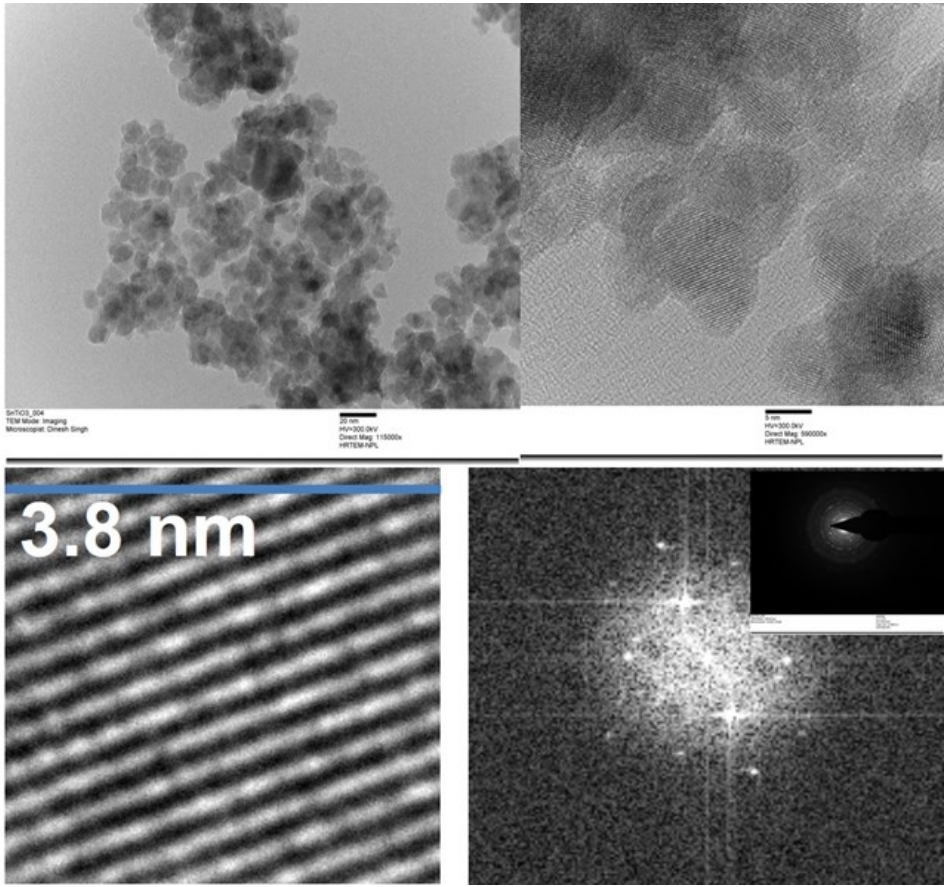

Fig. 10 depicts the high resolution TEM image of SnTiO3 nanocrystals.

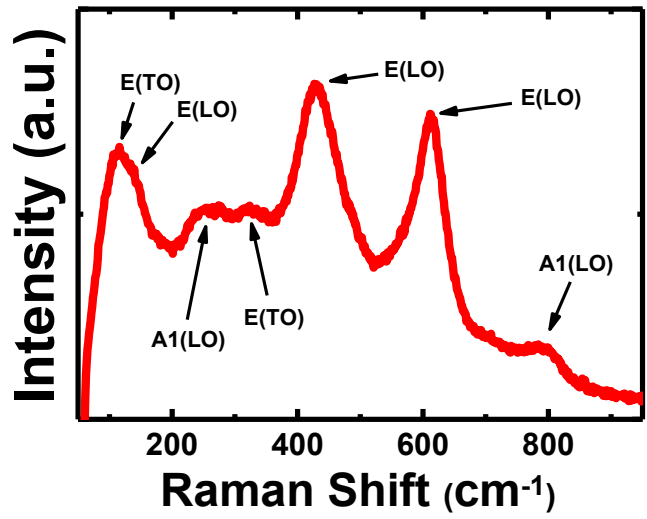

Fig. 11 Raman spectrum of tin titanate sample synthesised via sol gel route.

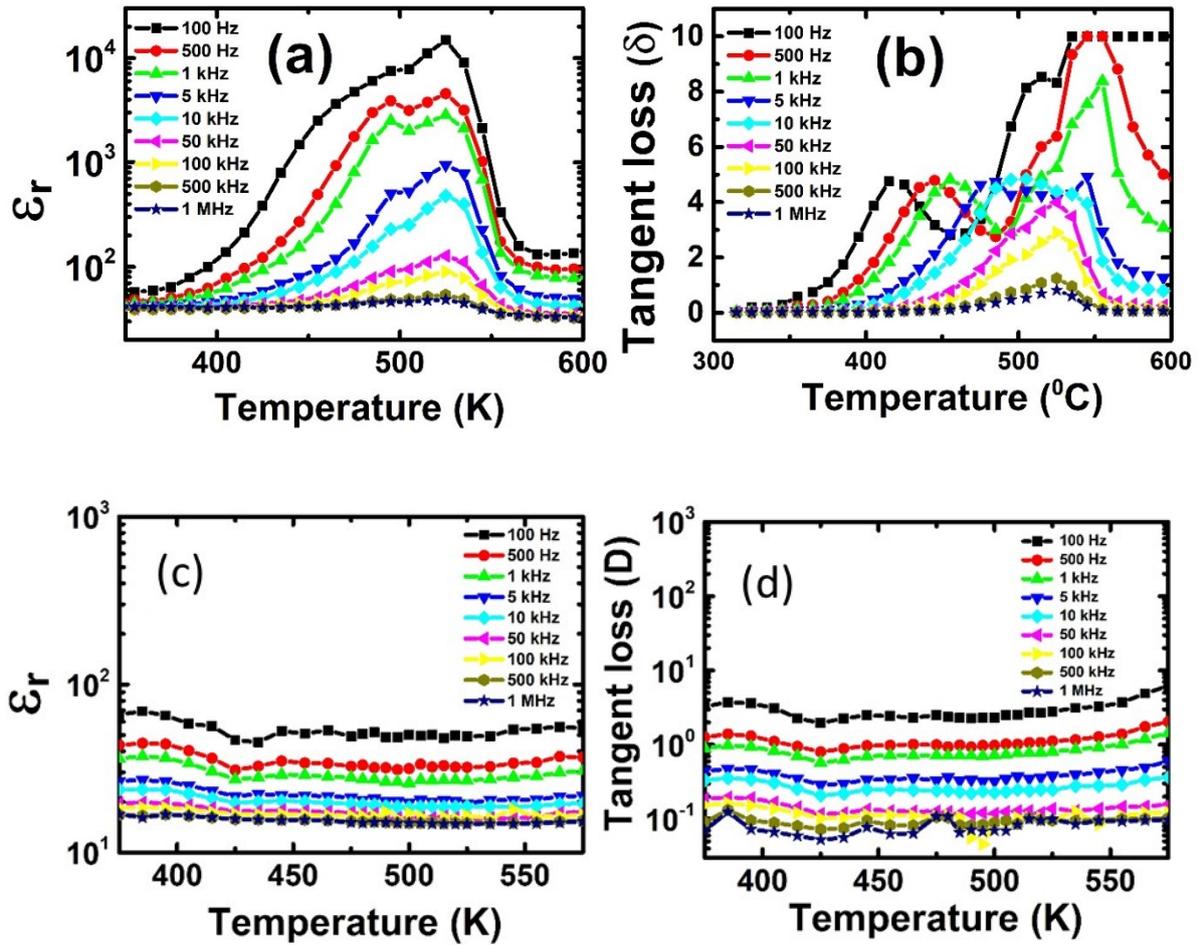

Figure 12: Relative permittivity, $\varepsilon_r$ and normalised loss, tan $\delta$, for tin titanate sol-gel pellets with Ag electrodes for: (a) and (b) initial heating and (c) and (d) subsequent cooling showing diminished peak.

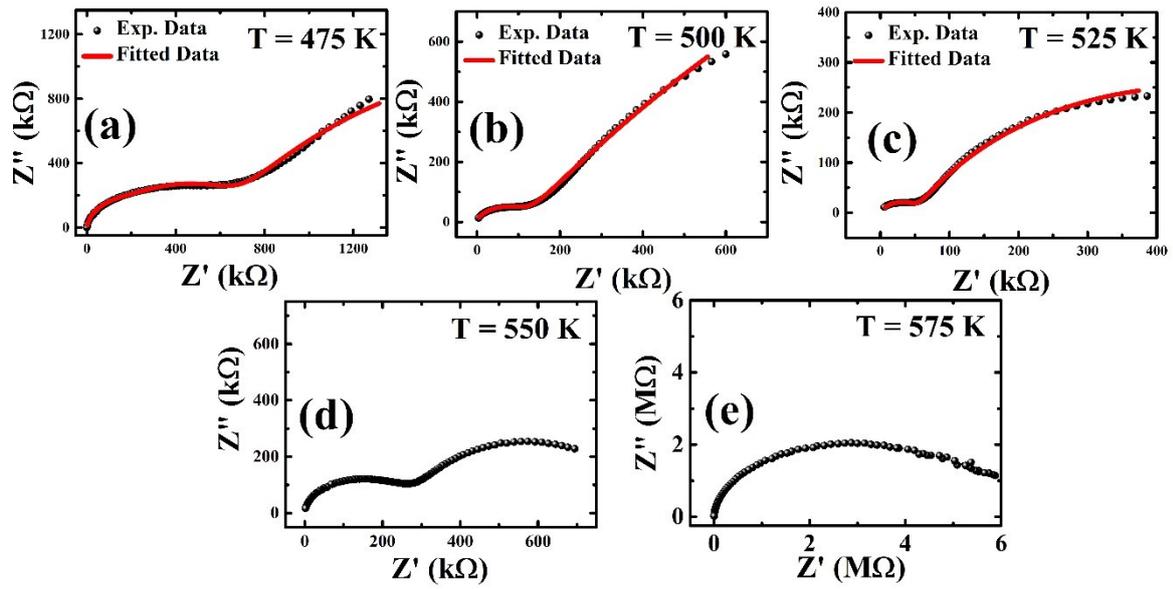

Figure 13: Impedance complex plane plots measured from tin titanate sol gel sample at different temperatures with equivalent electrical circuit model fitting.

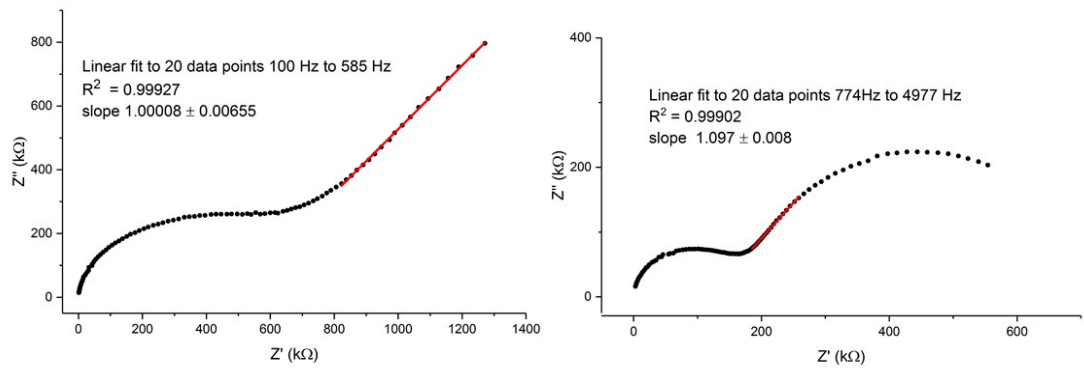

Figure 14: linear fit to low frequency data indicating Warburg element behaviour (gradient of fitted line ≈ 1) at 475 K and (right) apparent linearity present at higher temperatures (gradient of fitted line ≈ 1.1) at 540 K.

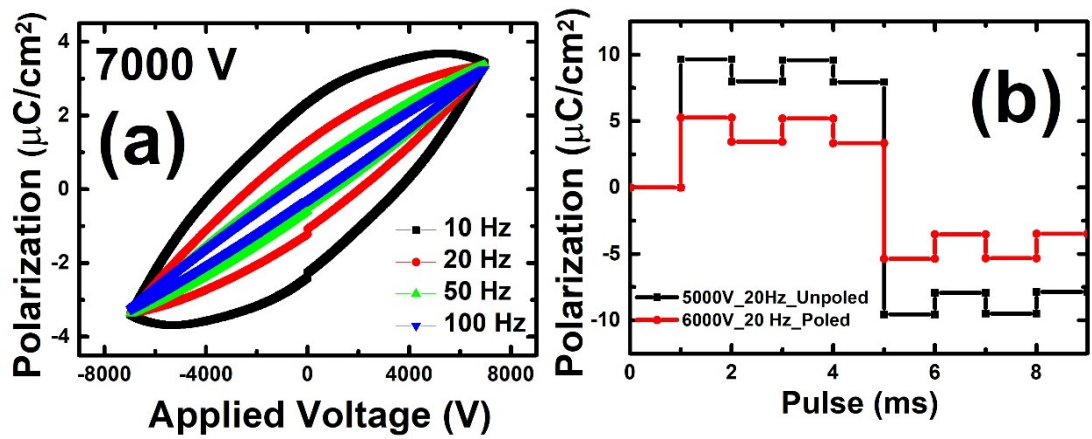

Figure 15: Polarisation vs. applied voltage plot of the tin titanate sol gel sample, (b) PUND data of the bulk sample in unpoled and poled situation. (a) Polarisation-voltage plot of tin titanate sample and (b) PUND data of the bulk sample in unpoled and poled situation.